\documentclass[preprint]{elsarticle}
\newcommand{\beq}{\begin{equation}} \newcommand{\eeq}{\end{equation}}
\newcommand{\bea}{\begin{eqnarray}} \newcommand{\eea}{\end{eqnarray}}
\newcommand{\bear}{\begin{eqnarray*}} \newcommand{\eear}{\end{eqnarray*}}
\newcommand{\lb}{\label} 
\newcommand{\rf}[1]{(\ref{#1})}   
\usepackage{epstopdf}
\usepackage{amsmath}
\usepackage{amsfonts}
\usepackage{amssymb}
\usepackage{makeidx}

\newtheorem{T}{Theorem}

\begin{document}

\title {Gauge Invariant Fractional Electromagnetic Fields.}

\author[imef]{Matheus Jatkoske Lazo}
\ead{matheuslazo@furg.br}
\address[imef]{Instituto de Matem\'atica, Estat\'\i stica e F\'\i sica - FURG, Rio Grande, RS, Brazil.}

\begin{abstract}

Fractional derivatives and integrations of non-integers orders was introduced more than three centuries ago but only recently gained more attention due to its application on nonlocal phenomenas. In this context, several formulations of fractional electromagnetic fields was proposed, but all these theories suffer from the absence of an effective fractional vector calculus, and in general are non-causal or spatially asymmetric. In order to deal with these difficulties, we propose a spatially symmetric and causal gauge invariant fractional electromagnetic field from a Lagrangian formulation. From our fractional Maxwell's fields arose a definition for the fractional gradient, divergent and curl operators.

\end{abstract}

\maketitle

\noindent


\section{Introduction}

The calculus with fractional derivatives and integrations of non-integers orders started more than three centuries ago with l'H\^opital and Leibniz when the derivative of order $\frac{1}{2}$ was suggested \cite{OldhamSpanier}. This subject was also considered by several mathematicians as Euler, Fourier, Liouville, Grunwald, Letnikov, Riemann and others up to nowadays. Although the fractional calculus is almost as old as the usual integer order calculus, only in the last three decades it has gained more attention due to its applications in various fields of science, engineering, economics, biomechanics, etc (see \cite{SATM,Kilbas,Hilfer,Magin} for a review).

Fractional derivatives are nonlocal operators and are historically applied in the study of nonlocal or time dependent processes. The first and well established application of fractional calculus in Physics was in the framework of anomalous diffusion, which are related to features observed in many physical systems (e.g. in dispersive transport in amorphous semiconductor, liquid crystals, polymers, proteins, etc \cite{Metzler,Metzler2,Klages}). Recently, the stud of nonlocal quantum phenomena through fractional calculus is in fast developing, where the nonlocal effects are due to either long-range interactions or time-dependent processes with many scales \cite{Hilfer,WBG,KBD,Laskin,Naber,Iomin,Tarasov1}. Relativistic quantum mechanics \cite{KP,Raspini,Zavada,MAB}, field theories \cite{Tarasov2,BGGB,LM,Goldfain,Herrmann} and gravitation \cite{Munkhammar} has been also recently considered in the context of fractional calculus. One of the most remarkable applications of fractional calculus in Physics was in the context of classical mechanics. Riewe \cite{Riewe} showed that Lagrangian involving fractional time derivatives leads to equation of motion with nonconservative forces such as friction. It is a remarkable result since frictional and nonconservative forces are beyond the usual macroscopic variational treatment \cite{Bauer}, and consequently, beyond the most advanced methods of classical mechanics. Riewe generalized the usual variational calculus for Lagrangian dependents on fractional derivatives \cite{Riewe} in order to deal with usual nonconservative forces. Recently, several approaches have been developed to generalize the least action principle and the Euler-Lagrange equations to includes fractional derivatives \cite{Riewe,Agrawal,BA,Cresson}. In this formalism, since the fractional derivatives are nonlocal, the Euler-Lagrange equations appear to not respect the causality principle when the Lagrangian contain fractional time derivatives. This difficulty is recently under investigation by several authors \cite{CI,DY}.

Although fractional calculus is an old subject and has several successful applications in Physics, the vector fractional calculus has only 10 years and all its different formulations have some problems of consistence (see \cite{Tarasov2} and references therein). Recently Tarasov \cite{Tarasov2} used the fractional generalization of the Fundamental Theorem of Calculus in order to fix some of these inconsistencies and formulate a fractional electromagnetic theory. Fractional Maxwell's equations can describe electromagnetic fields in media with fractional nonlocal properties, like in superconductor and semi-conductor physics \cite{Brandt,Belleguie,Genchev}, and in accelerated systems \cite{Mashhoon}. Several other fractional electromagnetic fields are proposed \cite{BGGB,Engheta,NaqviAbbas} by substituting the divergent and curl operators in the Maxwell's equations for fractional ones. As a consequence all these proposals for a fractional electromagnetic theory depend on the fractional vector calculus we choose. Furthermore, despite symmetries and gauge invariance have played a central role in theoretical physics, all these proposals also suffer from spatial asymmetries due to the fractional derivatives and are not gauge invariant. Recently, Herrmann \cite{Herrmann} has investigated gauge invariance in fractional fields in the context of fractional relativistic wave equations. He reproduced the spectrum of baryons accurately \cite{Herrmann}, stating fractional gauge fields as an interesting alternative approach to establish a $SU(n)$ symmetry in fields theories. Hence, a consistent formulation of fractional electromagnetic fields, as well as the investigation of gauge invariance, symmetries and causality, is important for several physical applications. In the present work, in order to not deal directly with the present difficulties of the fractional vector calculus, we take a different approach by starting with a formulating of a gauge invariant fractional field. We propose a fractional gauge invariant theory by introducing a fractional field strength tensor. The field's Lagrangian density is construct and the equations of motion are obtained from the fractional action principle \cite{Riewe,Agrawal,BA,Cresson}. From our generalized fractional Maxwell's fields arose a new definition for the fractional gradient, divergent and curl operators. However, we show that we can not obtain a causal gauge invariant fractional electromagnetic field with fractional time derivatives from a direct generalization of the field strength tensor and the action principle. A causal fractional theory is obtained only for first order time derivative and real order $\alpha >0$ spatial derivatives, resulting in a non-covariant theory.

Finally, in this paper we use the Riemann-Liouville fractional calculus \cite{OldhamSpanier}. It is the most popular fractional calculus approach. Although we choose the Riemann-Liouville calculus, our results are general and can be straightforward generalized to others fractional calculus, like the Caputo \cite{caputo} fractional calculus and others approaches \cite{OldhamSpanier,Kilbas,Podlubny}.

Our paper is organized as follows. In section 2 we review the Riemann-Liouville fractional calculus and we obtain the fractional Euler-Lagrange equation for a particular case of the more general fractional action principle \cite{Riewe,Agrawal,BA,Cresson}. A didactic discussion of the impossibility to obtain a causal electromagnetic field from a direct generalization of the field strength tensor and the action principle is done in section 3. In section 4, we propose a spatially symmetrical Lagrangian in order to obtain a gauge invariant fractional Maxwell's field. From our electromagnetic field arose a definition for the fractional vector operators. The conservation law for for electric charge and the fractional electromagnetic waves are studied in section 5. Finally, our conclusions are presented in Section 6.


\section{The Riemann-Liouville Fractional Calculus and Factional Action Principle}

Let $f:[a,b]\rightarrow \mathbb{R}$ be a real valued function $f\in C^{|n|}[a,b]$ where $n\in \mathbb{Z}$. We define the differ-integration operator $\!\!\!\!\!\phantom{D}_a D^n_x$ of integer order $n$ as
\beq
\lb{a1}
\!\!\!\!\!\phantom{D}_a D^n_x f(x) = \left\{ \begin{array}{ll} \frac{d^n}{dx^n}f(x) & (n>0) \\ f(x) & (n=0) \\ \int_{a}^x f(\tilde{x})(d\tilde{x})^{-n} & (n<0) \end{array} \right.
\eeq
where the $-n$-fold integration is defined as \cite{OldhamSpanier}
\bea
\lb{a2}
\int_{a}^x f(\tilde{x})(d\tilde{x})^{-n} & =& \int_{a}^x\int_{a}^{x_{-n}}\int_{a}^{x_{-n-1}}\cdots \int_{a}^{x_3}\int_{a}^{x_2} f(x_1)dx_1dx_2\cdots dx_{-n-1}dx_{-n}\nonumber \\
& =& \frac{1}{\Gamma(-n)}\int_{a}^x \frac{f(u)}{(x-u)^{1+n}}du \;\;\;\;\; (n\in \mathbb{Z_-}),
\eea
where the last equality follows from the Cauchy formula for repeated integration. 

In the Riemann-Liouville fractional calculus the left and the right fractional integral of order $\alpha,\beta \in \mathbb{R}$ are defined, respectively,  by the analytically continuation from \rf{a2}
\beq
\lb{a3}
\!\!\!\!\!\phantom{D}_a D^{\alpha}_x f(x) =\frac{1}{\Gamma(-\alpha)}\int_{a}^x \frac{f(u)}{(x-u)^{1+\alpha}}du \;\;\;\;\; (\alpha<0,\; \alpha,a\in \mathbb{R})
\eeq
and
\beq
\lb{a4}
\!\!\!\!\!\phantom{D}_x D^{\beta}_b f(x) =\frac{1}{\Gamma(-\beta)}\int_x^b \frac{f(u)}{(u-x)^{1+\beta}}du \;\;\;\;\; (\beta<0,\; \beta,b\in \mathbb{R}).
\eeq
The left and the right Riemann-Liouville fractional derivative of orders $\alpha,\beta >0$ ($\alpha,\beta\in \mathbb{R}$) are defined, respectively, by $\!\!\!\!\!\phantom{D}_a D^{\alpha}_x f(x) = \!\!\!\!\!\phantom{D}_a D^{\alpha-n}_x\!\!\!\!\!\phantom{D}_a D^{n}_x f(x)$ and $\!\!\!\!\!\phantom{D}_x D^{\beta}_b f(x)=(-1)^n\!\!\!\!\!\phantom{D}_x D^{\beta-n}_b \!\!\!\!\!\phantom{D}_x D^{n}_b f(x)$, namely
\beq
\lb{a5}
\!\!\!\!\!\phantom{D}_a D^{\alpha}_x f(x)=\frac{1}{\Gamma(n-\alpha)}\frac{d^n}{dx^n}\int_{a}^x \frac{f(u)}{(x-u)^{1+\alpha-n}}du \;\;\;\;\; (n=[\alpha]+1,\; \alpha\in \mathbb{R}_+^*, \; a\in \mathbb{R})
\eeq
and
\beq
\lb{a6}
\!\!\!\!\!\phantom{D}_x D^{\beta}_b f(x)=\frac{(-1)^n}{\Gamma(n-\beta)}\frac{d^n}{dx^n}\int_{x}^b \frac{f(u)}{(u-x)^{1+\beta-n}}du \;\;\;\;\; (n=[\beta]+1,\; \beta\in \mathbb{R}_+^*, \; b\in \mathbb{R}).
\eeq
where $\frac{d^n}{dx^n}$ stands for ordinary derivatives of integer order $n=[\alpha]+1$ or $n=[\beta]+1$. When $\alpha$ or $\beta$ is an integer, the Riemann-Liouville fractional derivative \rf{a5} and \rf{a6} reduces to ordinary derivatives. Finally, it is important to notice that the Riemann-Liouville differ-integration operator are nonlocal operators. The left (right) differ-integration operator \rf{a5} (\rf{a6}) depends on the values of the function at left (right) of $x$, i.e. $a\leq u \leq x$ ($x\leq u \leq b$).

Let us consider now a real valued function $f:\Omega \rightarrow \mathbb{R}$ of $d+1$ real variables $x^0,x^1,...x^d$ defined over the domain $\Omega = [a_0,b_0]\times \cdots \times [a_d,b_d]\subset \mathbb{R}^{d+1}$. We can define the left and the right Riemann-Liouville partial fractional derivatives of order $\alpha_{\mu},\beta_{\mu}\in \mathbb{R}_+^*$ with respect to $x^{\mu}$ as \cite{MAB} ($n_{\mu}=[\alpha_{\mu}]+1$ or $n_{\mu}=[\beta_{\mu}]+1$, $a_{\mu},b_{\mu}\in \mathbb{R}$) as
\beq
\lb{a7}
\!\!\!\!\!\phantom{D}_{a_{\mu}}\! \partial^{\alpha_{\mu}}_{\mu} f(x^0,...,x^d)=\frac{1}{\Gamma(n_{\mu}-\alpha_{\mu})}\partial_{x^{\mu}}^{n_{\mu}}\int_{a_{\mu}}^x \frac{f(x^0,...,x^{\mu-1},u,x^{\mu+1},...,x^d)}{(x^{\mu}-u)^{1+\alpha_{\mu}-n_{\mu}}}du
\eeq
and
\beq
\lb{a8}
\!\!\!\!\!\phantom{D}_{\mu} \partial^{\beta_{\mu}}_{b_{\mu}} f(x^0,...,x^d)=\frac{(-1)^{n_{\mu}}}{\Gamma(n_{\mu}-\beta_{\mu})}\partial_{x^{\mu}}^{n_{\mu}}\int_x^{b_{\mu}} \frac{f(x^0,...,x^{\mu-1},u,x^{\mu+1},...,x^d)}{(u-x^{\mu})^{1+\beta_{\mu}-n_{\mu}}}du,
\eeq
where $\partial_x^n$ is the ordinary partial derivative of integer order $n$ with respect to the variable $x$. If we identify the variable $x_0$ with the time, the left (right) differ-integration operator \rf{a7} (\rf{a8}) are dependent on the past (future). It is important to mention that if we want to construct a causal theory, we need to consider equations of motions for the fields with only left Riemann-Liouville time derivatives \rf{a7}.

For the generalized fractional electromagnetic field we proposed in section 4, it is convenient to introduced the left-right fractional Riemann-Liouville operators
\beq
\lb{b2b}
\partial_{\mu}^{\alpha\beta}=\frac{1}{2}\left(\!\!\!\!\!\phantom{D}_{a_{\mu}}\! \partial^{\alpha_{\mu}}_{\mu}-\!\!\!\!\!\phantom{D}_{\mu} \partial^{\beta_{\mu}}_{b_{\mu}}\right),
\eeq
since we are going to deal with Lagrangians ${\cal L} \left(A_{\mu},\frac{1}{2}\left(\!\!\!\!\!\phantom{D}_{a_{\mu}}\! \partial^{\alpha_{\mu}}_{\mu}-\!\!\!\!\!\phantom{D}_{\mu} \partial^{\beta_{\mu}}_{b_{\mu}}\right) A_{\mu},x^{\mu}\right)$ dependents on $N$ fields $A_{\mu}\equiv A_{\mu}(x^0,...x^d)$ ($\mu=1,2,...,N$) and its left-right Riemann-Liouville fractional derivatives of orders $0<\alpha_{\mu},\beta_{\mu}<1$. The minus signal in \rf{b2b} is due to the minus signal in \rf{a8} for $0<\beta_{\mu}<1$. It is important to notice that for $\alpha_{\mu}=\beta_{\mu}$ the operators \rf{b2b} becomes Riesz fractional derivatives \cite{Agrawal2,Podlubny}. Furthermore, we want to find the extremum condition for fractional actions $S$ defined by
\beq
\lb{a9}
S=\int_{\Omega} {\cal L} (A_{\mu},\partial_{\nu}^{\alpha\beta} A_{\mu},x^{\mu})(dx_{\mu}),
\eeq
where we follows the notation in line with \cite{MAB}, where ${\cal L} (A_{\mu},\cdots)$ $\equiv$ ${\cal L} (A_0,...A_{N},\cdots)$, ${\cal L} (\cdots,\partial_{\nu}^{\alpha\beta} A_{\mu},\cdots)$ $\equiv$ ${\cal L} (\cdots,\partial_{0}^{\alpha\beta} A_{0},...,\partial_{d}^{\alpha\beta} A_{0},...,\partial_{0}^{\alpha\beta} A_{N},...,\partial_{d}^{\alpha\beta} A_{N}\cdots)$, ${\cal L} (\cdots,x^{\mu})$ $\equiv$ ${\cal L} (\cdots,x^{0},...,x^{d})$ and $(dx^{\mu})$ $\equiv$ $dx^0\cdots dx^d$. In order to develop the action principle of our model, it is convenient to formulate an fractional Euler-Lagrange equation for \rf{a9}. We introduce the following theorem:
\begin{T}
Let $\alpha_{\mu},\beta_{\mu}\in \mathbb{R}$ with $0<\alpha_{\mu},\beta_{\mu}<1$, and $S$ be an action of the form
\beq
\lb{t1}
S=\int_{\Omega} {\cal L} \left(A_{\mu},\frac{1}{2}\left(\!\!\!\!\!\phantom{D}_{a_{\nu}}\! \partial^{\alpha_{\nu}}_{\nu}-\!\!\!\!\!\phantom{D}_{\nu} \partial^{\beta_{\nu}}_{b_{\nu}}\right) A_{\mu},x^{\mu}\right)(dx_{\mu}),
\eeq
defined on a set of $N$ fields  $A_{\mu} \in C^1[a,b]$ and satisfying the boundary conditions $A_{\mu}(a)=A_{a \mu}$ and $A_{\mu}(b)=A_{b \mu}$. Also let ${\cal L}\in C^{2}[a,b]\times \mathbb{R}^{2N}$. Then the necessary condition for $S$ to possess an extremum is that the fields $A_{\mu}$ fulfills the following fractional Euler-Lagrange equation:
\beq
\lb{t2}
\frac{\partial {\cal L}}{\partial A_{\mu}}-\partial_{\nu}^{\beta\alpha}\frac{\partial {\cal L}}{\partial(\partial_{\nu}^{\alpha\beta} A_{\mu})}=0,
\eeq 
where we follows the standard convention where we perform a summation over repeated indices.
\end{T}
{\it proof.} In order to develop the necessary conditions for the extremum of the action \rf{t1}, we define a family of fields $A_{\mu}$
\beq
\lb{p1}
A_{\mu}=A_{\mu}^*+\varepsilon \eta_{\mu},
\eeq
where $A_{\mu}^*$ is the desired real function that satisfy the extremum of \rf{t1}, $\varepsilon \in \mathbb{R}$, and $\eta_{\mu}\in C^1[a,b]$ satisfy the boundary conditions
\beq
\lb{p2}
\eta_{\mu}(a)=\eta_{\mu}(b)=0.
\eeq
The condition for the extremum is obtained when the first G\^ateaux variation is zero:
\beq
\lb{p3}
\begin{split}
\delta S&=\lim_{\varepsilon \rightarrow 0} \frac{S[A_{\mu}^*+\varepsilon \eta_{\mu}]-S[A_{\mu}^*]}{\varepsilon}\\
&=\int_{\Omega}\left( \eta_{\mu}\frac{\partial{\cal L}}{\partial A_{\mu}^*} +\frac{\!\!\!\!\!\phantom{D}_{a_\nu}\! \partial^{\alpha_{\nu}}_{\nu}\eta_{\mu}}{2}\frac{\partial{\cal L}}{\partial (\partial_{\nu}^{\alpha\beta}A_{\mu}^*)} -\frac{\!\!\!\!\!\phantom{D}_{\nu} \partial^{\beta_{\nu}}_{b_{\nu}}\eta_{\mu}}{2}\frac{\partial{\cal L}}{\partial (\partial_{\nu}^{\alpha\beta}A_{\mu}^*)}\right)(dx_{\mu})=0.
\end{split}
\eeq
Using the formula for integration by part \cite{OldhamSpanier,SATM}, the boundary conditions \rf{p2} and the fundamental lemma of the calculus of variations, we obtain the fraction Euler-Lagrange equations \rf{t2}. Finally, it is important to mention that our action principle generalizes \cite{Agrawal2} and is a particular case of the more general theorem proposed in \cite{Agrawal}.


\section{A Non-Causal Gauge Invariant Fractional Maxwell Field}

The main objective of this section is show that it is not possible to obtain a causal fractional electromagnetic field with fractional time derivatives from a direct generalization of the field strength tensor and the action principle, by substituting the integer order derivatives by fractional ones. By causal fractional electromagnetic fields we means a theory described by fractional Maxwell's equations containing only left Riemann-Liouville time derivatives. As we stated in the last section, left Riemann-Liouville time derivatives are nonlocal operators dependent on the past time, consequently, they are nonlocal causal operators. On the other hand, if right Riemann-Liouville time derivatives are present in the fractional Maxwell's equations, the fields will be dependent on the future time, resulting in a non-causal theory. 

Let $A_{\mu}=(\psi,-\mathbf{A})$ be the $4$-vector electromagnetic potential. The usual Maxwell's field strength tensor $F_{\mu \nu}$ is given by
\beq
\lb{b1}
F_{\mu \nu}=\partial_{\mu}A_{\nu}-\partial_{\nu}A_{\mu},
\eeq
where we use, as usual, $x_0\equiv ct$, and the Minkowski metric $\eta^{\mu \nu}=\eta_{\mu \nu}={\mbox{diag}}(+1,-1,-1,-1)$, and we use the standard convention that the Latin indices run only over the space coordinates (i.e. $i,j,k,...=1,2,3$) and the Greek indices (excluding $\alpha,\beta$) includes both time and space coordinates (i.e. $\mu,\nu,\sigma,...=0,1,2,3$). The direct fractional generalization of \rf{b1} we propose is obtained by changing the order of the differentiation $\partial_{\mu}$ from the integer order $1$ to an arbitrary order $0<\alpha\leq 1$, where $\alpha\in \mathbb{R}$, namely
\beq
\lb{b2}
F_{\mu \nu}^{\alpha}=\!\!\!\!\!\phantom{D}_{\mu} \partial^{\alpha}_{b}A_{\nu}-\!\!\!\!\!\phantom{D}_{\nu} \partial^{\alpha}_{b}A_{\mu},
\eeq
where the components of the fractional field strength tensor $F_{\mu \nu}^{\alpha}$, defined over the domain $\Omega = [a_0,b_0]\times [a_1,b_1] \times [a_2,b_2] \times [a_3,b_3]\subset \mathbb{R}^{4}$, are identified with the fractional electromagnetic fields by
\beq
\lb{b3}
F_{\mu \nu}^{\alpha}=\left(\begin{array}{cccc}
0 & E^{\alpha}_x & E^{\alpha}_y & E^{\alpha}_z \\ 
-E^{\alpha}_x & 0 & -B^{\alpha}_z & B^{\alpha}_y \\ 
-E^{\alpha}_y & B^{\alpha}_z & 0 & -B^{\alpha}_x \\ 
-E^{\alpha}_z & -B^{\alpha}_y & B^{\alpha}_x & 0
\end{array} \right),
\eeq
where we have defined the $F_{0 i}^{\alpha}$ components to be the fractional electric fields and the $F_{i j}^{\alpha}$ components to be the fractional magnetic fields. The fractional electromagnetic fields defined by \rf{b2} and \rf{b3} are nonlocal fields due to the right Riemann-Liouville partial fractional derivatives in \rf{b2}. We choose the non-causal right Riemann-Liouville derivatives in \rf{b2} in order to obtain a causal second pair of fractional Maxwell's equations and non-causal first pair ones, as we will show in this section. On the other hand, if we choose left Riemann-Liouville derivatives on \rf{b2}, it results in a non-causal second pair of fractional Maxwell's equations and causal first pair ones.

It is important to notice that, like in the standard electromagnetic theory,  all the physical properties of the fractional electromagnetic field are determined not by the potential $A_{\mu}$, but rather by the tensor $F_{\mu \nu}^{\alpha}$. The reason for this is that fractional electromagnetic tensor \rf{b2} exhibits gauge invariance. By changing the $4$-vector potentials in the following way
\beq
\lb{b4}
A_{\mu}\longrightarrow A_{\mu}+\!\!\!\!\!\phantom{D}_{\mu} \partial^{\alpha}_{b}\phi,
\eeq
the effect we have on the fractional field strength tensor is
\beq
\lb{b5}
\delta F_{\mu \nu}^{\alpha}=\!\!\!\!\!\phantom{D}_{\mu} \partial^{\alpha}_{b}(A_{\nu}+\!\!\!\!\!\phantom{D}_{\nu} \partial^{\alpha}_{b}\phi)-\!\!\!\!\!\phantom{D}_{\nu} \partial^{\alpha}_{b}(A_{\mu}+\!\!\!\!\!\phantom{D}_{\mu} \partial^{\alpha}_{b}\phi)-F_{\mu \nu}^{\alpha}=\!\!\!\!\!\phantom{D}_{\mu} \partial^{\alpha}_{b}\!\!\!\!\!\phantom{D}_{\nu} \partial^{\alpha}_{b}\phi-\!\!\!\!\!\phantom{D}_{\nu} \partial^{\alpha}_{b}\!\!\!\!\!\phantom{D}_{\mu} \partial^{\alpha}_{b}\phi=0.
\eeq
Thus, the transformation \rf{b4} does not change the form of the fractional field strength tensor. The gauge invariance of the tensor $F_{\mu \nu}^{\alpha}$ \rf{b2} will play a fundamental role in the formulation of a causal second pair of fractional Maxwell's equations. From the fractional Euler-Lagrange equations \cite{Agrawal} it follows that tensor fields defined as in \rf{b2} (dependent only on right fractional derivatives) will result in fractional equations of motions dependents only on left fractional derivatives. As a consequence, these equations of motions for the electromagnetic fields \rf{b2} are causal.

Finally, the Lagrangian density of the fractional electromagnetic field can be construct from the tensor $F_{\mu \nu}^{\alpha}$ defined in \rf{b2} by
\beq
\lb{b6}
{\cal L}_{Field}=-\frac{1}{16\pi c}F_{\mu \nu}^{\alpha} F_{\alpha}^{\mu \nu},
\eeq
where $F_{\alpha}^{\mu \nu}=\eta^{\mu \rho} \eta^{\nu \sigma}F_{\rho \sigma}^{\alpha}$. The fractional electromagnetic action is then given by
\beq
\lb{b7}
S=-\frac{1}{16\pi c}\int_{\Omega} F_{\mu \nu}^{\alpha} F_{\alpha}^{\mu \nu} (dx_{\mu})-\frac{1}{c^2}\int_{\Omega}j^{\mu}A_{\mu} (dx_{\mu}),
\eeq
where $(dx_{\mu})=cdtd^3x$, $j^{\mu}=(c\rho,\mathbf{j})$ is the $4$-vector current and the second term on the right hand side accounts the interaction between the matter and the fields, and the domain $\Omega = [a_0,b_0]\times [a_1,b_1] \times [a_2,b_2] \times [a_3,b_3]\subset \mathbb{R}^{4}$ with $-\infty <a_{\mu} <b_{\mu}< \infty$ is the time-space physical dimension of the system. The necessary condition for extremum of the action functional defined above is given by the fractional Euler-Lagrange equations \cite{Agrawal}. From \cite{Agrawal} we obtain the fractional generalization of the second pair of Maxwell's equations
\beq
\lb{b8}
\!\!\!\!\!\phantom{D}_{a} \partial^{\alpha}_{\nu}F_{\alpha}^{\mu \nu}=-\frac{4\pi}{c}j^{\mu}.
\eeq
These equations are causal in the sense we have only left fractional derivatives on \rf{b8}. This fact is more evident if we identify the respective components of the field strength tensor with the electric and magnetic fields. From \rf{b8} we can rewrite the second pair of fractional Maxwell's equations \rf{b8} in a more familiar form
\beq
\lb{b9}
\begin{split}
\mbox{div}_{-}^{\alpha} \mathbf{E}^{\alpha}&=4\pi \rho \\
\mbox{curl}_{-}^{\alpha}\mathbf{B}^{\alpha}&=\frac{4\pi}{c}\mathbf{j}+\frac{1}{c^{\alpha}}\!\!\!\!\!\phantom{D}_{a} \partial^{\alpha}_{t}\mathbf{E}^{\alpha},
\end{split}
\eeq
where $\!\!\!\!\!\phantom{D}_{a} \partial^{\alpha}_{t}\equiv c^{\alpha}\!\!\!\!\!\phantom{D}_{a} \partial^{\alpha}_{0}$, and we defined the left fractional divergent and curl of a vector field $\mathbf{F}$ as
\beq
\lb{b10}
\mbox{div}_{-}^{\alpha} \mathbf{F} \equiv \!\!\!\!\!\phantom{D}_{a} \partial^{\alpha}_{i}F_i \;\;\; \mbox{and}\;\;\; \mbox{curl}_{-}^{\alpha} \mathbf{F} \equiv \mathbf{e}_i\varepsilon_{ijk}\!\!\!\!\!\phantom{D}_{a} \partial^{\alpha}_{j}F_k,
\eeq
where $\mathbf{e}_i$ are orthogonal unit vectors, $F_i$ are components of the vector field $\mathbf{F}=F_i\mathbf{e}_i$ and $\varepsilon_{ijk}$ is the Levi-Civita symbol. In \rf{b9} we have only one time derivative and it is a left fractional derivative, resulting in a causal second pair of fractional Maxwell's equations. However, an additional non-causal field equation should be added in order to completely fix the fractional electromagnetic fields. Like in the usual electromagnetic fields, it is important to notice that the second pair of fractional Maxwell's equations do not completely fix the properties of the fields. It can be seen from the fact that \rf{b9} fix the time variation of the fractional electric field (related to the fractional derivative $\!\!\!\!\!\phantom{D}_{a} \partial^{\alpha}_{t}\mathbf{E}^{\alpha}$) but not the time variations of the magnetic field. In order to completely fix the fields, we need to obtain a fractional generalization of the first pair of Maxwell's equations. From \rf{b2} it is easy to check that
\beq
\lb{b11}
\!\!\!\!\!\phantom{D}_{\rho} \partial^{\alpha}_{b}F_{\mu \nu}^{\alpha}+\!\!\!\!\!\phantom{D}_{\mu} \partial^{\alpha}_{b}F_{\nu \rho}^{\alpha}+\!\!\!\!\!\phantom{D}_{\nu} \partial^{\alpha}_{b}F_{\rho \mu}^{\alpha}=0.
\eeq
By identifying the components of \rf{b11} with the electric and magnetic fields we obtain the non-causal first pair of fractional Maxwell's equations
\beq
\lb{b12}
\begin{split}
\mbox{div}_{+}^{\alpha} \mathbf{B}^{\alpha}&=0 \\
\mbox{curl}_{+}^{\alpha}\mathbf{E}^{\alpha}&=-\frac{1}{c^{\alpha}}\!\!\!\!\!\phantom{D}_{t} \partial^{\alpha}_{b}\mathbf{B}^{\alpha},
\end{split}
\eeq
where
\beq
\lb{b13}
\mbox{div}_{+}^{\alpha} \mathbf{F} \equiv \!\!\!\!\!\phantom{D}_{i} \partial^{\alpha}_{b}F_i \;\;\; \mbox{and}\;\;\; \mbox{curl}_{+}^{\alpha} \mathbf{F} \equiv \mathbf{e}_i\varepsilon_{ijk}\!\!\!\!\!\phantom{D}_{j} \partial^{\alpha}_{b}F_k
\eeq
are right fractional divergent and curl operators. The left \rf{b10} and right \rf{b13} divergent and curl operators that arouse in our Maxwell's equations are in line with the fractional operators defined in the Tarasov fractional vector calculus \cite{Tarasov2}. Despite the non-causality of the fractional electromagnetic fields defined from \rf{b1}, we can make another criticism to the fractional Maxwell's fields studied in this section. While the first pair of fractional Maxwell's equations \rf{b12} contain only right Riemann-Liouville fractional derivatives, the second pair ones \rf{b9} contain only left derivatives. On other words, the differential equations on \rf{b9} depends on values of the fields at the left of the spatial coordinates $x_i$ and the differentials equations on \rf{b12} depends on values at the right of the spacial coordinates. There is no physical justification for this spacial asymmetry of the Maxwell's equations. Furthermore, due to this asymmetry, a continuity equations for the sources can not be obtained from the fractional Maxwell's fields \rf{b9} and \rf{b12}. 


\section{A Generalized Gauge Invariant Fractional Electromagnetic Field}

In order to deal with the spacial asymmetry of the Maxwell's equations \rf{b12} and \rf{b9} and obtain a spacial symmetrical theory, in this section we propose a fractional generalization of \rf{b2} containing both left and right fractional derivatives. Furthermore, the fractional Maxwell's equations we obtain give us a new definition on fractional divergent and curl operators. The generalization of \rf{b2} we propose is given by
\beq
\lb{c2}
F_{\mu \nu}^{\alpha\beta}=\frac{1}{2}(\!\!\!\!\!\phantom{D}_{a_{\mu}}\! \partial^{\alpha_{\mu}}_{\mu}-\!\!\!\!\!\phantom{D}_{\mu} \partial^{\beta_{\mu}}_{b_{\mu}})A_{\nu}-\frac{1}{2}(\!\!\!\!\!\phantom{D}_{a_\nu}\! \partial^{\alpha_{\nu}}_{\nu}-\!\!\!\!\!\phantom{D}_{\nu} \partial^{\beta_{\nu}}_{b_{\nu}})A_{\mu}=\partial_{\mu}^{\alpha\beta}A_{\nu}-\partial_{\nu}^{\alpha\beta}A_{\mu},
\eeq
where $\alpha_{\mu},\beta_{\mu}\in \mathbb{R}$ with $0<\alpha_{\mu},\beta_{\mu}<1$, and we introduced the left-right Riemann-Liouville fractional derivative \rf{b2b}. In \rf{c2} we consider the most general case where to each component $x^{\mu}$ we associated a different fractional Riemann-Liouville derivative (of order $0<\alpha_{\mu}\leq 1$ and dependent of the real parameter $a_{\mu}$ if a left derivative, and of order $0<\beta_{\mu}\leq 1$ and dependent of the parameter $b_{\mu}$ if right ones). The resulting Maxwell's equations will display spacial symmetry if the two order parameters are equal $\alpha_{\mu}=\beta_{\mu}$, and if $b_{\mu}=-a_{\mu}$. Finally, the components of the two parameters fractional field strength tensor $F_{\mu \nu}^{\alpha\beta}$ are identified with the fractional electromagnetic fields as in \rf{b3}.

Like in the previous section, it is easy to check that the fields \rf{c2} are gauge invariant. By change the $4$-vector potential in the following way
\beq
\lb{c4}
A_{\mu}\longrightarrow A_{\mu}+\partial_{\mu}^{\alpha\beta}\phi,
\eeq
the effect they have on the tensor of the electromagnetic field is null, 
\beq
\lb{c5}
\begin{split}
\delta F_{\mu \nu}^{\alpha\beta}&=\partial_{\mu}^{\alpha\beta}(A_{\nu}+\partial_{\nu}^{\alpha\beta}\phi)-\partial_{\nu}^{\alpha\beta}(A_{\mu}+\partial_{\mu}^{\alpha\beta}\phi)-F_{\mu \nu}^{\alpha\beta}\\
&=\partial_{\mu}^{\alpha\beta}\partial_{\nu}^{\alpha\beta}\phi -\partial_{\nu}^{\alpha\beta}\partial_{\mu}^{\alpha\beta}\phi =0.
\end{split}
\eeq
It is important to mention that, since our field strength tensor is gauge invariant, we can formulate a fractional generalization for the Lorenz gauge condition. By taking the left-right Riemann-Liouville fractional derivative of the $4$-vector potential and equating to zero, we obtain the fractional Lorenz gauge condition:
\beq
\lb{c5b}
\partial_{\mu}^{\alpha\beta}A^{\mu}=\partial_{i}^{\alpha\beta}A_{i}+\partial_{0}^{\alpha\beta}\psi=0,
\eeq
where $A^{\mu}=\eta^{\mu\nu}A_{\nu}$.

The action for the fractional generalized electromagnetic field is defined as in \rf{b6} as
\beq
\lb{c6}
S=-\frac{1}{16\pi c}\int_{\Omega} F_{\mu \nu}^{\alpha\beta} F_{\alpha\beta}^{\mu \nu} (dx_{\mu})-\frac{1}{c^2}\int_{\Omega}j^{\mu}A_{\mu} (dx_{\mu}),
\eeq
where $F_{\alpha\beta}^{\mu \nu}=\eta^{\mu \rho} \eta^{\nu \sigma}F_{\rho \sigma}^{\alpha\beta}$. It is important to notice that the action \rf{c6} is a functional on the left-right Riemann-Liouville derivatives \rf{b2b} of the potential instead of the right Riemann-Liouville derivative as in \rf{b7}. The necessary condition for extremum of the action \rf{c6} is given by the fractional Euler-Lagrange equations \rf{t2}. It give us the fractional generalization of the second pair of Maxwell's equations
\beq
\lb{c7}
\partial_{\nu}^{\beta \alpha}F_{\alpha\beta}^{\mu \nu}=\frac{4\pi}{c}j^{\mu},
\eeq
where $\partial_{\mu}^{\beta \alpha}$ are the fractional left-right Riemann-Liouville derivative operators \rf{b2b}. It is important to remember that in \rf{c7} the Einstein summation is realized only for index $\mu,\nu,\rho$, etc but not for the fractional index $\alpha$ and $\beta$. As in the previous section, in order to completely fix the fractional electric and magnetic fields, we need to generalize the first pair of Maxwell's equations. From \rf{c2}, it is easy to check that
\beq
\lb{c9}
\partial_{\rho}^{\alpha\beta}F_{\mu \nu}^{\alpha\beta}+\partial_{\mu}^{\alpha\beta}F_{\nu \rho}^{\alpha\beta}+\partial_{\nu}^{\alpha\beta}F_{\rho \mu}^{\alpha\beta}=0.
\eeq
By introducing the left-right fractional gradient, divergent and curl operators, for a field $\mathbf{F}$ and a function $\phi$, as
\beq
\lb{c11}
\mbox{grad}^{\alpha\beta} \phi \equiv \mathbf{e}_i\partial_{i}^{\alpha\beta}\phi, \;\;\;\mbox{div}^{\alpha\beta} \mathbf{F} \equiv \partial_{i}^{\alpha\beta}F_i \;\;\; \mbox{and}\;\;\; \mbox{curl}^{\alpha\beta} \mathbf{F} \equiv \mathbf{e}_i\varepsilon_{ijk}\partial_{j}^{\alpha\beta}F_k,
\eeq
and by identifying the components of \rf{c7} and \rf{c9} with the electric and magnetic fields as defined in \rf{b3} (with $\mathbf{E}^{\alpha\beta}$ and $\mathbf{B}^{\alpha\beta}$ instead of $\mathbf{E}^{\alpha}$ and $\mathbf{B}^{\alpha}$, respectively) we obtain now the following fractional Maxwell's equations
\beq
\lb{c10}
\begin{split}
\mbox{div}^{\beta\alpha}\mathbf{E}^{\alpha\beta}&=4\pi \rho \\
\mbox{curl}^{\beta\alpha}\mathbf{B}^{\alpha\beta}&=\frac{4\pi}{c}\mathbf{j}+\partial_{0}^{\beta\alpha}\mathbf{E}^{\alpha\beta},\\
\mbox{div}^{\alpha\beta} \mathbf{B}^{\alpha\beta}&=0, \\
\mbox{curl}^{\alpha\beta}\mathbf{E}^{\alpha\beta}&=-\partial_{0}^{\alpha\beta}\mathbf{B}^{\alpha\beta}.
\end{split}
\eeq
The field equations \rf{c10} contain now both left and right Riemann-Liouville derivatives. By choosing $\alpha_{\mu}=\beta_{\mu}$ and $b_{\mu}=-a_{\mu}$ the left-right operators reduces to Riesz derivatives and the fractional Maxwell's equations becomes spatially and time symmetric. Furthermore, when $\alpha_0=\beta_0=1$, the fractional time derivatives on the left-hand side of \rf{c10} reduces to usual first order derivatives. In this case the electromagnetic fields resulting from \rf{c10} are causal. Finally, it is important to notice that we do not started with a previous definition of fractional vector calculus operators. It is from our electromagnetic fields that arose a new definition for the fractional divergent and curl operators \rf{c9}. In order to complete our fractional vector calculus, we also add a definition for a fractional gradient operator. A criticism to our approach is that, different from \cite{Tarasov2}, we can not formulate fractional generalization of integral vector theorems. As a consequence, it is not possible to obtain the Maxwell's equations in integral form for our fractional gauge invariant fields.


\section{Charge conservation and wave equations}

In this section we investigate the charge conservation and wave equations for spatially symmetrical and causal gauge fields. For this case ($\alpha_{\mu}=\beta_{\mu}$ and $\alpha_0=\beta_0=1$) the time and space derivatives in \rf{c10} becomes integer order time and fractional order Riesz derivatives, respectively. The gauge invariance and the presence of Riesz derivatives also display a important role in the electric charge conservation law and in the existence of fractional waves. By taking the left-right divergent on the second equation of \rf{c10} and using the first one we obtain the fractional continuity equation
\beq
\lb{c11}
\mbox{div}^{\alpha} \mathbf{j} +\frac{1}{c}\partial_{t}^{2} \rho=0,
\eeq
where the Riez divergent operator $\mbox{div}^{\alpha}\equiv \mbox{div}^{\alpha\alpha}$ is defined by \rf{c11} with $\alpha_{\mu}=\beta_{\mu}$. Equation \rf{c11} is a fractional generalization of the continuity equations and is the differential law for the charge conservation in fractional nonlocal electrodynamics. On the other hand, for a source free field, we obtain the following fractional wave equations for the electric and magnetic fields
\beq
\lb{c12}
\begin{split}
&\frac{1}{c^2}\partial_{t}^2\mathbf{E}^{\alpha\alpha}-\left(\mbox{div}^{\alpha}\right)^2\mathbf{E}^{\alpha\alpha}=0,\\
&\frac{1}{c^2}\partial_{t}^2\mathbf{B}^{\alpha\alpha}-\left(\mbox{div}^{\alpha}\right)^2\mathbf{B}^{\alpha\alpha}=0,
\end{split}
\eeq
where we take the time derivative of the second and fourth equations in \rf{c10} with $\alpha_{\mu}=\beta_{\mu}$ and $\alpha_0=\beta_0=1$. Furthermore, due to the gauge invariance of our fields, we also can obtain fractional wave equations for the vector and scalar potentials. By inserting $\mathbf{E}^{\alpha\alpha}=-\mbox{grad}^{\alpha}\psi-\frac{1}{c}\partial_{t}\mathbf{A}$ and $\mathbf{B}^{\alpha\alpha}=\mbox{curl}^{\alpha}\mathbf{A}$ into the second pair of Maxwell's equations \rf{c10}, and by using the fractional Lorenz gauge \rf{c5b} we obtain:
\beq
\lb{c13}
\begin{split}
&\frac{1}{c^2}\partial_{t}^2\mathbf{A}-\left(\mbox{div}^{\alpha}\right)^2\mathbf{A}=0,\\
&\frac{1}{c^2}\partial_{t}^2\psi-\left(\mbox{div}^{\alpha}\right)^2\psi=0,
\end{split}
\eeq
where $\mbox{grad}^{\alpha}\equiv \mbox{grad}^{\alpha\alpha}$ and $\mbox{curl}^{\alpha}\equiv \mbox{curl}^{\alpha\alpha}$ are the Riez gradient and curl operator defined in \rf{c11} with $\alpha_{\mu}=\beta_{\mu}$, respectively.

Finally, let us consider only the plane-wave solutions for spatially symmetrical and causal fields. In this case the wave equations \rf{c12} and \rf{c13} reduces to a fractional partial differential equation of the form
\beq
\lb{c14}
\frac{1}{c^2}\partial^2_t u(x,t)-\left(\partial_{x}^{\alpha}\right)^2 u(x,t)=0,
\eeq
where for simplicity we consider only $x$-dependence, and $\partial_{x}^{\alpha}\equiv \partial_{x}^{\alpha\alpha}$ is a Riesz derivative. In order to derive the plane-wave solution for fractional Maxwell's fields we consider $x \in \mathbb{R}$ ($a_{i}=-b_{i}=\infty$), $t\in \mathbb{R}^+$ ($a_0=0$ and $b_0=\infty$) and smooth space periodic $u(x,t)$ functions. By closely following the method developed in \cite{ZhangLiu} we expand $u(x,t)$ in Fourier series:
\beq
\lb{c15}
u(x,t)=\sum_{k=-\infty}^{\infty}f_k(t)e^{ikx},
\eeq
where $f_k(t)$ are smooth functions. Inserting \rf{c15} into the wave equation \rf{c14} and using the proposition $2.1$ in \cite{ZhangLiu} we obtain the following ordinary differential equation for the functions $f_k(t)$:
\beq
\lb{c16}
\ddot f_k(t)=-|k|^{2\alpha}c^2\sin^2\left(\frac{\alpha \pi}{2}\right)f_k(t).
\eeq
From \rf{c16} we obtain the plane-wave solution for the causal and spatially symmetrical fractional gauge fields introduced in the previous section:
\beq
\lb{c17}
u(x,t)=\sum_{k=-\infty}^{\infty}f_{1,k}\cos(\omega_{\alpha}t)e^{ikx}+\sum_{k=-\infty,k\neq 0}^{\infty}f_{2,k}\omega_{\alpha}^{-1}\sin(\omega_{\alpha}t)e^{ikx},
\eeq
where the constants $f_{1,k}$ and $f_{2,k}$ are the coefficients of Fourier series expansions of the initial value conditions $u(x,0)$ and $\partial_t u(x,0)$, respectively, and
\beq
\lb{c18}
\omega_{\alpha}=|k|^{\alpha}c\sin\left(\frac{\alpha \pi}{2}\right)
\eeq
is the fractional time frequency of plane-waves. In the limit $\alpha \rightarrow 1$ \rf{c18} reduces to the usual dispersion relation relating time and space frequencies. Hence, equation \rf{c18} give us an easy experimental test for nonlocality effects in fractional electrodynamics.


\section{Conclusion}

In this work we introduced a fractional gauge invariant electromagnetic field from a Lagrangian formulation. Different from other proposals for a fractional Maxwell's field \cite{Tarasov2,BGGB,Engheta,NaqviAbbas} our model is gauge invariant and display spatial symmetry. Furthermore, while all previous proposals suffer from the absence of an effective fractional vector calculus, from our generalized fractional Maxwell's equations arose a definition for the fractional gradient, divergent and curl operators. We also show that we can not obtain a causal gauge invariant fractional electromagnetic field with fractional time derivatives from a direct
generalization of the field strength tensor and the action principle. A causal fractional theory is obtained only for first order time derivative and arbitrary real order spatial derivatives. In the last section we investigate the charge conservation law and wave equations for our model. We obtained the plane-wave solution that give us an easy experimental test for our model. Finally, the action method we developed in this paper is general and can be used to construct others fractional field theories.


\section*{Acknowledgements}
This work has been supported by CNPq and CAPES (Brazilian agencies).


\end{document}